\pdfoutput=1
\documentclass[a4paper,11pt]{article}
\usepackage{pos}
\usepackage{pgfplots}
\hypersetup{pdftitle={Machine learning for the LHC physics program: a 2025-2026 stocktake}, pdfauthor={Jesse Thaler}}
\pgfplotsset{compat=1.18}
\definecolor{posblue}{HTML}{1F4E8C}

\DeclareRobustCommand{\Sec}[1]{Sec.~\ref{#1}}

\DeclareRobustCommand{\Fig}[1]{Fig.~\ref{#1}}

\DeclareRobustCommand{\Appx}[1]{Appendix~\ref{#1}}

\DeclareRobustCommand{\Reference}[1]{Ref.~\cite{#1}}

\renewcommand{\logo}{\relax}  

\title{Machine learning for the LHC physics program:  \\ a 2025--2026 stocktake}
\ShortTitle{Machine learning for the LHC physics program: a 2025--2026 stocktake}

\author*{Jesse Thaler}

\affiliation{Center for Theoretical Physics -- a Leinweber Institute, Massachusetts Institute of Technology, \\
Cambridge, Massachusetts, United States}
\affiliation{Institut des Hautes \'Etudes Scientifiques, 91440 Bures-sur-Yvette, France}
\affiliation{Institut de Physique Th\'eorique, CEA Paris-Saclay, 91191 Gif-sur-Yvette, France}
\affiliation{The NSF Institute for Artificial Intelligence and Fundamental Interactions}

\emailAdd{jthaler@mit.edu}

\abstract{The first sentence of this abstract---and the introduction to these proceedings---was authored by a human, but the bulk of this document was generated by an agentic AI system.  In this talk, I take stock of machine learning (ML) for the LHC physics program over the twelve months from May 2025 to May 2026.  The corpus is the HEPML Living Review, split at May 2025 into 1{,}756 earlier papers and 569 later ones.  An AI pipeline surveyed the 569 abstracts, ranked them by citations, recency, theme, and collaboration involvement, and read 103 papers in full (95 from after the split, plus 8 earlier baseline papers), producing a structured note for each.  The notes were then synthesized into six claims about the state of the field, checked by independent reviewer agents, and re-verified against the source papers.  The headline claim is that (1) ML for high-energy physics (HEPML) stopped being a research area that builds tools and became infrastructure that the LHC physics program depends on: ATLAS and CMS now publish physics results that depend on neural networks, and the archived ALEPH data have re-entered production.  The other five claims are: (2) simulation-based inference and foundation models are two revolutions starting to merge; (3) AI agents are the genuinely new front, with 47 papers tagged as agents or LLMs in twelve months and no adopted measurement yet; (4) ``do we trust it?'' is the fastest-growing agenda, with one recent paper in five about uncertainty, calibration, or interpretability; (5) what is slowing down is informative, since equivariance was absorbed into a tool and model-specific phenomenology ceded ground to model-agnostic searches; and (6) theory ML crossed a capability threshold in multiple research areas.  I close with what is settled, what is incoming, and what is open, and briefly discuss the concerns raised by this way of working with AI.}

\FullConference{14th Edition of the Large Hadron Collider Physics (LHCP2026)\\
18-22 May 2026\\
Paris, France\\}

\tableofcontents  

\begin{document}
\maketitle

\newpage

\section{Introduction}

When I agreed to give an LHCP plenary talk back in November 2025, I knew I wanted to talk about the exciting ways that artificial intelligence (AI) was changing how particle physicists were collecting, analyzing, and interpreting data from the Large Hadron Collider (LHC).
What I hadn't anticipated, though, was that AI was poised to evolve dramatically in the subsequent months.
By May 2026, the relevant question was no longer whether AI was having an impact at the LHC (it was), but rather whether AI was going to fundamentally change the scientific enterprise (it will).

What happened between November 2025 and May 2026?
For one thing, commercial AI tools from frontier labs gained a surprising proficiency in science-related tasks.
I saw this first-hand on sabbatical in Paris during the 2025--26 academic year, when I split my time between Institut des Hautes \'Etudes Scientifiques (IHES), Institut de Physique Th\'eorique (IPhT), and the myriad cafés and libraries in Paris itself.
Armed with a trial subscription to Claude Max 20x, I used the expansive time of sabbatical to try out agentic AI coding.
Working with Claude Opus 4.1 was incredibly frustrating, especially all of its hallucinated Mathematica commands.
Instead of learning how to vibe code, I ended up coding laboriously by hand, which eventually turned into \Reference{Benjamin:2026lbj} (unrelated to the LHC).
While \Reference{Benjamin:2026lbj} does credit Opus 4.1 for one key scientific insight, AI had not (yet) lived up to its hype.

Then, Opus 4.5 was released on November 24, 2025.
Suddenly, I was working with an AI system that could actually code proficiently, including finding Mathematica commands not in the standard documentation.
Around the same time, I was finalizing a report on \href{https://arxiv.org/abs/2509.02661}{``The Future of Artificial Intelligence and the Mathematical and Physical Sciences (AI+MPS)''}~\cite{Ferguson:2026aimps}, which had a short section on the potential of AI co-pilots.
Seemingly overnight, that potential had turned into a reality.
While I was not ready to write an entirely AI-generated research paper (cf.~\cite{Schwartz:2026ekw}), I realized that I needed to switch my attitude from principled skeptic to inquisitive explorer.

How far could I push agentic AI systems?
I launched my first audacious experiment on February 20, 2026 at 2pm (Paris time):  two minutes of prompting yielded a 30-minute talk on \href{https://lfs.jthaler.net/talks/jthaler_2026_03_24_AnnualReviews_Salon.pdf}{``Predictably Uncertain: Academic Publishing in the Era of AI,''} which I presented at an Annual Reviews salon in March 2026 to a trepidatious audience.
This was admittedly a bit of a parlor trick, intended to illustrate AI's capabilities more than illuminate AI's consequences.
But I was (and still am) convinced that a key barrier to responsible AI use is simply not knowing what is possible.

During the long weekend of May 8--11, 2026, I saturated my (now paid) Max 20x plan to create an LHCP plenary slide deck based on \href{https://github.com/iml-wg/HEPML-LivingReview}{``A Living Review of Machine Learning for Particle Physics''}~\cite{Feickert:2021ajf,krause_2026_21626667} and 103 selected papers from it~\cite{Birk:2024knn, Mikuni:2024qsr, Huetsch:2024quz, Favaro:2024rle, ATLAS:2024xxl, Krause:2024avx, Brehmer:2024yqw, CMS:2024nsz, ATLAS:2025kuz, Benato:2025rgo, ATLAS:2025rbr, Janssen:2025zke, ATLAS:2025dkv, Spinner:2025prg, Puljak:2025gtj, Benevedes:2025nzr, Cheng:2025ewj, Astrand:2025sij, Bothmann:2025lwg, Araz:2025oax, Buss:2025cyw, Ghosh:2025fma, Sun:2025orx, Falcao:2025jom, Canelli:2025ybb, Costantini:2025wxp, Vent:2025ddm, Park:2025ebs, Favaro:2025pgz, Que:2025mhf, Bahl:2025xvx, Villadamigo:2025our, Raikwar:2025fky, Diefenbacher:2025zzn, Chu:2025jsi, Hallin:2025ywf, CMS:2025dsh, Bakshi:2025fgx, CMS:2025kje, Fernando:2025xzv, Electron-PositronAlliance:2025hze, Bhimji:2025isp, Laatu:2025xsw, Bonanno:2025pdp, Vega:2025hgz, Frank:2025zuk, Gendreau-Distler:2025fsj, Menzo:2025cim, Vega:2025xma, Mikuni:2025ocp, Bahl:2026qaf, Zoccheddu:2026cjr, Nguyen:2026wsv, Emami:2026zhi, Elsharkawy:2026kwp, Bahl:2026jvt, LHCb:2026ezb, Hsu:2026sww, Zheng:2026juf, Plehn:2026gxv, Bellscheidt:2026rjh, Esmail:2026jpb, Heimel:2026hgp, Cruz-Martinez:2026rct, Alharazin:2026lcb, Reuter:2026pdv, Valsecchi:2026kpp, Hill:2026naa, Haide:2026bmu, Govorkova:2026wqp, Vigl:2026ppx, Khanpour:2026erj, Aarrestad:2026xrs, Knipfer:2026kng, Akar:2026yjn, Pfahler:2026fpj, Ore:2026qgp, Atif:2026eju, Badea:2026klb, Defranchis:2026wyw, Patel:2026npj, Li:2026krn, ClarkeHall:2026ewp, Mallampalli:2026hrl, Shih:2026lmy, Conde:2026csr, Woodward:2026jnx, ATLAS:2026ovz, Moreno:2026mqk, Agrawal:2026lvg, Mikuni:2026ced, Li:2026azw, Auvinen:2026haf, Addepalli:2026hwb, MODE:2026xoy, Menzo:2026qrl, Cardona-Giraldo:2026rnn, Shih:2026jfe, Sadek:2026fkr, Alonso-Monsalve:2026okj, CMS:2026uph, Krzmanc:2026fdw, Barrue:2026qul}, followed by multiple rounds of human and AI feedback.
While it would have been faster (and arguably more fun) for me to write an LHCP talk in my own words (and with my own graphic design), the AI analysis went far beyond anything that I could reasonably do as a human, even if I might quibble with some of the AI's conclusions (and cringe at its stylistic choices).
As a finishing touch, I added a human-authored preamble (not reproduced in these proceedings) and two meta-commentary codas (I leave it to Claude to decide if they earn a place here).
I then braced myself for the ire of my colleagues when I delivered \href{https://indico.cern.ch/event/1537360/contributions/6833617/}{``my'' talk} on May 20, 2026.
Quoting from slide 11:
\begin{quote}
\emph{[Insert AI-generated content here]}
\end{quote}


\section{About this talk: a deliberate use of AI}
\label{sec:about}

Before the physics, I should disclose how the slides I presented at LHCP were made: Claude Opus 4.7 produced them under my direction, from the selection of papers and the reading to the structured notes, the drafting, and the fact-checking.
The process took roughly 72 hours of compute and went through multiple drafting passes, independent reviewer agents, and a verification pass over the numeric claims.
The talk was mine; the work was the AI's.
This write-up was made the same way, by Claude Fable 5.1 working from the slides and the pipeline's notes.
Every number below was checked against its source paper, not the pipeline's notes, by a separate verification pass.
Where a number differs from the slides shown on May~20, I follow the source paper.

The concerns about AI in scientific research are real.
Authorship norms in academia do not yet accommodate semi-autonomous AI agents; closed-source commercial tools are acting as research infrastructure; fact-checking AI output at scale is hard for humans; and reproducibility across model versions is not guaranteed.
The benefits are also real: 103 papers read in full, beyond any single human's bandwidth; automated cross-checking by independent verifier passes; constraints like research-group balance and citation honesty enforced systematically; and conflicts of interest disclosed because the pipeline was told to.
The sharpest conflict is that five of the papers I discuss below~\cite{Moreno:2026mqk, Badea:2026klb, Menzo:2026qrl, Hill:2026naa, Shih:2026lmy}, and at least four more of the 103 papers read~\cite{Shih:2026jfe, Gendreau-Distler:2025fsj, Plehn:2026gxv, Knipfer:2026kng}, were themselves produced with Anthropic models, so an Anthropic model is reporting here on work done with Anthropic models.
Apply the appropriate discount.

\section{How this talk was put together}
\label{sec:method}

Every conference plenary talk begins with the obligatory apology:  ``Because of limited time, I can only present a subset of interesting results.''
This talk operationalized the apology.
The corpus is the \href{https://github.com/iml-wg/HEPML-LivingReview}{HEPML Living Review}~\cite{Feickert:2021ajf,krause_2026_21626667}, which indexed 2{,}325 papers when the snapshot was taken on May~8, 2026 (\Reference{krause_2026_21626667} is the archived June 2026 release).
One rule splits it into two eras: a paper is ``post'' if its arXiv identifier carries \texttt{YYMM}~$\geq$~\texttt{2505}, and ``pre'' otherwise.
That isolates the 569 papers from May 2025 through April 2026 against a baseline of 1{,}756.
May 2025 is also the training cutoff of the Claude model that built the slides, so the era split doubles as the line between what the model might already have seen and what it had to read fresh.
That is convenient, and it is also a confound: the model knows the post-split papers from reading them and the pre-split ones from memory, and that asymmetry favors a before-and-after story.

\paragraph{Selection and reading.}
Themes were assigned by keyword rules on titles and abstracts, with a paper allowed to carry several tags.
The post-split papers were ranked by \href{https://inspirehep.net/}{INSPIRE} citation count, citations per month, a theme weight, and a major-collaboration flag, with 2026 papers reweighted upward to offset citation-age bias.
The top 95 of that ranking~\cite{ATLAS:2025kuz, Benato:2025rgo, ATLAS:2025rbr, Janssen:2025zke, ATLAS:2025dkv, Spinner:2025prg, Puljak:2025gtj, Benevedes:2025nzr, Cheng:2025ewj, Astrand:2025sij, Bothmann:2025lwg, Araz:2025oax, Buss:2025cyw, Ghosh:2025fma, Sun:2025orx, Falcao:2025jom, Canelli:2025ybb, Costantini:2025wxp, Vent:2025ddm, Park:2025ebs, Favaro:2025pgz, Que:2025mhf, Bahl:2025xvx, Villadamigo:2025our, Raikwar:2025fky, Diefenbacher:2025zzn, Chu:2025jsi, Hallin:2025ywf, CMS:2025dsh, Bakshi:2025fgx, CMS:2025kje, Fernando:2025xzv, Electron-PositronAlliance:2025hze, Bhimji:2025isp, Laatu:2025xsw, Bonanno:2025pdp, Vega:2025hgz, Frank:2025zuk, Gendreau-Distler:2025fsj, Menzo:2025cim, Vega:2025xma, Mikuni:2025ocp, Bahl:2026qaf, Zoccheddu:2026cjr, Nguyen:2026wsv, Emami:2026zhi, Elsharkawy:2026kwp, Bahl:2026jvt, LHCb:2026ezb, Hsu:2026sww, Zheng:2026juf, Plehn:2026gxv, Bellscheidt:2026rjh, Esmail:2026jpb, Heimel:2026hgp, Cruz-Martinez:2026rct, Alharazin:2026lcb, Reuter:2026pdv, Valsecchi:2026kpp, Hill:2026naa, Haide:2026bmu, Govorkova:2026wqp, Vigl:2026ppx, Khanpour:2026erj, Aarrestad:2026xrs, Knipfer:2026kng, Akar:2026yjn, Pfahler:2026fpj, Ore:2026qgp, Atif:2026eju, Badea:2026klb, Defranchis:2026wyw, Patel:2026npj, Li:2026krn, ClarkeHall:2026ewp, Mallampalli:2026hrl, Shih:2026lmy, Conde:2026csr, Woodward:2026jnx, ATLAS:2026ovz, Moreno:2026mqk, Agrawal:2026lvg, Mikuni:2026ced, Li:2026azw, Auvinen:2026haf, Addepalli:2026hwb, MODE:2026xoy, Menzo:2026qrl, Cardona-Giraldo:2026rnn, Shih:2026jfe, Sadek:2026fkr, Alonso-Monsalve:2026okj, CMS:2026uph, Krzmanc:2026fdw, Barrue:2026qul}, plus 8 pre-split anchors for the baseline of \Sec{sec:baseline}~\cite{Birk:2024knn, Mikuni:2024qsr, Huetsch:2024quz, Favaro:2024rle, ATLAS:2024xxl, Krause:2024avx, Brehmer:2024yqw, CMS:2024nsz}, were read in full: 103 papers, each yielding one structured note in a fixed schema (claim, method, dataset, one headline number, why it matters for the LHC) and one candidate figure.
Twenty-eight of them made it onto the main slides~\cite{Birk:2024knn, Mikuni:2024qsr, Huetsch:2024quz, Krause:2024avx, Brehmer:2024yqw, CMS:2024nsz, ATLAS:2025kuz, ATLAS:2025rbr, ATLAS:2025dkv, Spinner:2025prg, Bahl:2025xvx, Villadamigo:2025our, Diefenbacher:2025zzn, Chu:2025jsi, CMS:2025dsh, Electron-PositronAlliance:2025hze, Bhimji:2025isp, Bonanno:2025pdp, Bahl:2026qaf, Hill:2026naa, Mallampalli:2026hrl, Shih:2026lmy, Moreno:2026mqk, Mikuni:2026ced, Menzo:2026qrl, Shih:2026jfe, Krzmanc:2026fdw, Barrue:2026qul}.
Two evidential levels follow: every share, count, and lift below comes from all 569 post-split abstracts; every named paper comes from the 103 read in full.
Abstracts were harvested only from 2024 onward, so theme shares compare the 569 post-split papers with the 514 papers of the preceding sixteen months (January 2024 through April 2025).
Below, the arXiv-class comparison uses the full pre-split baseline, and the rate comparison uses calendar 2024, the last full year before the split.

\begin{figure}[t]
\centering
\begin{minipage}[b]{0.555\textwidth}\centering
\begin{tikzpicture}
\begin{axis}[width=\linewidth, height=3.6cm, ybar, bar width=1.7pt, ymin=0, ymax=88,
  xmin=0.3, xmax=64.7, xtick={1,13,25,37,49,61}, xticklabels={2021,2022,2023,2024,2025,2026},
  x tick label style={anchor=north west, xshift=-2pt}, xtick align=outside, ytick align=outside,
  ylabel={papers per month}, ylabel near ticks, ylabel style={font=\scriptsize},
  tick label style={font=\scriptsize}, axis lines=left, axis line style={-}, clip=false,
  every axis plot/.append style={fill=posblue, draw=none}]
\addplot[fill=posblue, draw=none] coordinates {(1,15) (2,14) (3,29) (4,17) (5,23) (6,21) (7,21) (8,13) (9,26) (10,27) (11,19) (12,24) (13,13) (14,11) (15,35) (16,13) (17,11) (18,11) (19,17) (20,14) (21,7) (22,20) (23,54) (24,36) (25,25) (26,26) (27,36) (28,38) (29,39) (30,38) (31,27) (32,25) (33,30) (34,26) (35,33) (36,34) (37,43) (38,37) (39,34) (40,32) (41,35) (42,27) (43,37) (44,21) (45,23) (46,34) (47,35) (48,34) (49,22) (50,29) (51,37) (52,34) (53,29) (54,36) (55,43) (56,42) (57,80) (58,52) (59,49) (60,54) (61,42) (62,43) (63,67) (64,32)};
\draw[dashed, gray!80!black, line width=0.5pt] (axis cs:52.5,0) -- (axis cs:52.5,86) node[anchor=south, font=\scriptsize] {May 2025};
\end{axis}
\end{tikzpicture}
\end{minipage}\hfill
\begin{minipage}[b]{0.425\textwidth}\centering
\begin{tikzpicture}
\begin{axis}[width=0.84\linewidth, height=4.2cm, xbar, bar width=5pt, xmin=0, xmax=8.4, ymin=0.3, ymax=12.7,
  y dir=reverse, ytick={1,...,12}, yticklabels={{AI agents / LLMs},{Quantum ML},{Trigger / FPGA},{Uncertainty / calibration},{Track / vertex reco.},{SBI / unbinned},{Foundation models},{Theory ML},{Anomaly detection},{Generative simulation},{BSM phenomenology},{Equivariance}},
  y tick label style={font=\fontsize{6.5}{7}\selectfont}, xtick={0,2,4,6,8},
  xlabel={theme-share lift (post / pre)}, xlabel style={font=\scriptsize}, x tick label style={font=\scriptsize},
  xtick align=outside, ytick align=outside, axis lines=left, axis line style={-}, clip=false,
  nodes near coords, nodes near coords style={font=\fontsize{5.5}{6}\selectfont, text=black, /pgf/number format/fixed, /pgf/number format/precision=2, /pgf/number format/fixed zerofill},
  every axis plot/.append style={fill=posblue, draw=none}]
\addplot[fill=posblue, draw=none] coordinates {(7.08,1) (2.37,2) (2.31,3) (1.58,4) (1.52,5) (1.38,6) (1.34,7) (1.05,8) (1.03,9) (0.97,10) (0.81,11) (0.8,12)};
\draw[dashed, gray!80!black, line width=0.5pt] (axis cs:1,0.3) -- (axis cs:1,12.7);
\end{axis}
\end{tikzpicture}
\end{minipage}
\caption{\emph{Left:} monthly publication rate in the HEPML Living Review, January 2021 through April 2026, with the May 2025 era split marked; the post-split rate averages 47 papers per month, against 33 in 2024. \emph{Right:} theme-share lift across the split (post-split share divided by pre-split share) for 12 of the 21 themes, including the highest and lowest; the dashed line marks equal shares (lift 1).}
\label{fig:method}
\end{figure}
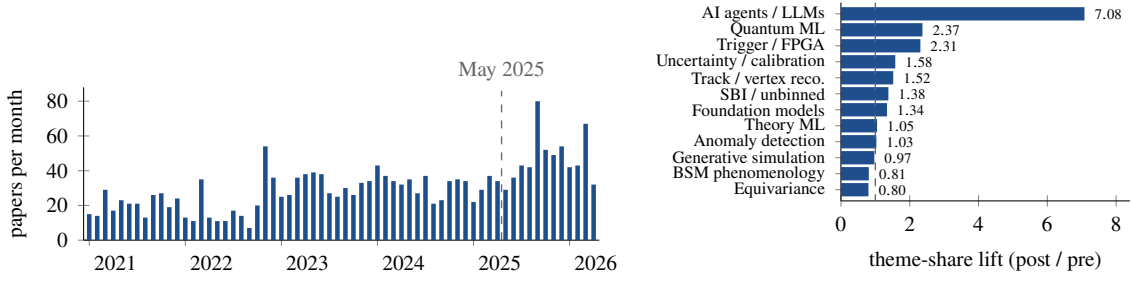

\paragraph{A field that grew up while you weren't looking.}
The publication rate (\Fig{fig:method}) rose by roughly 45\%, from about 33 papers a month in 2024 to an average of 47 since the split, with the rise concentrated from September 2025 on.
The composition moved with it: the \texttt{hep-ex} share of arXiv primary classes rose from 21\% to 24\%, and the \texttt{cs.LG} share nearly doubled from 2.4\% to 4.4\%; the major collaborations (ATLAS, CMS, LHCb, and the groups working the ALEPH archive) account for about 8\% of recent output.
The field is modestly more experimental, and more tied to the machine-learning literature, than a year ago.

The statistic I lean on most is the \emph{share lift} of each theme: its post-split share of papers divided by its pre-split share.
There are two caveats: papers carry 1.9 tags each after the split against 1.7 before, so the neutral value is nearer 1.1 than 1.0, and the keyword rules were written after reading the post-split literature, which favors themes with new vocabulary.
Across the 21 named themes the lift ranges from 0.80 for equivariant networks to 7.08 for AI agents and large language models, the steepest rise by a wide margin; eight themes sit below 1.0, with equivariance and BSM phenomenology (0.81) lowest.
Volume tells a different story: the largest post-split themes are uncertainty and calibration (114 papers), theory ML (99), generative simulation (87), and jet tagging (82), while the steepest grower, AI agents and LLMs, has 47 papers.
Steepest is not biggest.
Two of the steepest risers in \Fig{fig:method}, quantum machine learning and trigger/FPGA inference, get no claim of their own here; the latter was the subject of \href{https://indico.cern.ch/event/1537360/contributions/6833618/}{the plenary immediately before this one}.

If your favorite paper is not here, the reason is arithmetic: 569 papers in the era, 95 of them read in full, and a selection that favors citations and 2026 freshness, hence established groups.
Forty-seven papers a month.
You can't keep up.
Neither can I.

\section{Where we stood at LHCP 2025}
\label{sec:baseline}

A year ago, the field was building tools.
Equivariant transformers (L-GATr) had been shown to scale, in time and memory, far beyond the thousand or so particles at which equivariant graph networks run out of memory, with Lorentz symmetry respected by construction~\cite{Brehmer:2024yqw}.
Foundation models were proposals rather than products: the first cross-task foundation model for particle physics~\cite{Birk:2024knn}, and an independent study solving three key collider-physics challenges~\cite{Mikuni:2024qsr}.
Generative simulation had matured into a community benchmark, with 31 submissions compared head to head on shower quality, generation time, and model size~\cite{Krause:2024avx}.
CMS's first model-agnostic resonance search had appeared, reporting improvements of up to $7.1\times$ in the expected cross-section limit and $6.4\times$ in the cross section needed for a $5\sigma$ discovery, for $G_{\mathrm{KK}}\to HH$ relative to an inclusive selection~\cite{CMS:2024nsz}.
ATLAS had published a simultaneous unbinned measurement of twenty-four $Z$+jets observables~\cite{ATLAS:2024xxl}, and a comprehensive comparison study had surveyed unbinned machine-learning unfolding as a coming paradigm, not yet a standard~\cite{Huetsch:2024quz}.
For the state of play a year ago, the two LHCP 2025 plenaries by \href{https://indico.cern.ch/event/1419878/contributions/6424738/}{D.~Guest} (experiment) and \href{https://indico.cern.ch/event/1419878/contributions/6424774/}{C.~Krause} (theory and analysis) remain the reference points.
Before May 2025, the tools were built.
After May 2025, they got used.

\section{Six claims about how the field changed for LHCP 2026}
\label{sec:claims}

Five years ago the question was ``can ML do physics?''
One year ago it was ``what tools should we build?''
Today, after twelve months in which 569 papers were indexed, I make six claims about how the field changed.

\subsection{Claim 1: from tools to infrastructure}
\label{sec:claim1}

HEPML transitioned from a research area that builds tools into infrastructure that the LHC physics program depends on: pre-split papers built methods; post-split papers deploy them.
ATLAS has published GN2, an end-to-end transformer flavor tagger deployed for Run~3.
It replaces the earlier DL1d, improves $c$-jet rejection by $3.5\times$ at 70\% $b$-tagging efficiency, and projects up to 30\% better HL-LHC sensitivity to Higgs pair production and the charm Yukawa coupling~\cite{ATLAS:2025dkv}.
More telling than a deployment is a physics result that depends on one.
CMS measured the charm and bottom Yukawa couplings simultaneously in $t\bar tH$ events and, together with its earlier $VH$ search, obtained a world-leading direct constraint, $|\kappa_c|<3.5$ (observed) at the 95\% confidence level---feasible only because two networks combine, ParticleNet for jet flavor identification and a Particle Transformer for event classification~\cite{CMS:2025dsh}.
Calibrating such classifiers has itself become an ML task~\cite{ATLAS:2025rbr}, and even legacy data have re-entered production, with archived ALEPH events supporting an unbinned thrust measurement~\cite{Electron-PositronAlliance:2025hze} and modern flavor tagging in hadronic $Z$ decays~\cite{Defranchis:2026wyw}.
The caveat: the five papers above are the whole of my evidence for a claim about the whole field, because the corpus was never classified by deployment.
Part of the effect is also publication cadence, since pipelines that were already live are only now being written up.
Even so, the CMS measurement makes the point: \emph{the physics result depends on the network.}

\subsection{Claim 2: two revolutions, converging}
\label{sec:claim2}

Two revolutions are under way, and they are starting to merge.
Simulation-based inference (SBI) is the methodological one: a histogram throws away information the moment it bins, and SBI works event by event over the full phase space instead.
The methodology is reaching the LHC and has already landed at LEP.
Neural SBI extracts the gluon parton distribution directly from the unbinned $t\bar t$ phase space, with uncertainties the authors find similar to NNPDF4.0's for $x$ between 0.01 and 0.35, and much less sensitive than the binned fit to the systematic uncertainties that double its error around $x \approx 0.1$.
It is a proof of concept on simulated data, not yet a measurement~\cite{Barrue:2026qul}.
At LEP, OmniFold applied to 1994 ALEPH $e^+e^-$ data reveals a small but systematic shift of the thrust distribution toward larger $\tau = 1-T$ relative to the original binned result, supports finer binning, and discriminates between the Pythia~8, Herwig~7, and Sherpa~2 parton showers~\cite{Electron-PositronAlliance:2025hze}.
In short, \emph{SBI changes how we report results: likelihoods rather than histograms.}

Foundation models are the architectural revolution, and they have gone from proposal to product.
OmniLearned, pretrained on more than one billion jets, sets the state of the art on the community top-tagging benchmark with a background rejection $1/\varepsilon_B = 3486\pm157$ at 30\% signal efficiency, fine-tunes to ATLAS full simulation, and transfers to CMS Open Data and to few-GeV MINERvA neutrino interactions~\cite{Bhimji:2025isp, Krzmanc:2026fdw}.
A dedicated review maps the landscape~\cite{Hallin:2025ywf}.
For now, the convergence is a direction rather than a settled fact, since most papers still pick one revolution.
The nearest concrete instance pairs a foundation model with a third agenda: it sits inside a resonant anomaly-detection analysis on real CMS Open Data, and its authors invite scrutiny and report that their background estimate does not yet describe the signal region~\cite{Mikuni:2026ced}.
In short, \emph{foundation models pretrain and fine-tune, and now comes the hard part, which is to measure.}

\subsection{Claim 3: AI agents are the genuinely new front}
\label{sec:claim3}

If one theme deserves the word ``new,'' it is this one.
Papers tagged ``AI agents / large language models'' went from 6 before the split to 47 after it, a share lift of 7.08, the steepest of any theme.
The jump is far outside counting fluctuations; the uncertainty is in the tag, which does not separate autonomous agents from other uses of language models.
Twelve months in, the cohort has already differentiated into four uses.
\textbf{Analysis automation:} multi-agent pipelines that execute end-to-end HEP measurements on open data with known answers.
One of them reproduces eight ALEPH, DELPHI, and CMS open-data analyses (two repeated on both LEP data sets): the $Z$ lineshape; $R_b$, $R_c$, and $A^{b}_{\mathrm{FB}}$; energy-energy correlators; the Lund jet plane; $N_\nu$ from $\Gamma_{\mathrm{inv}}$; and $H\to\tau\tau$.
It takes roughly 4--10 hours of wall-clock time per measurement, idle waits for usage-limit resets included~\cite{Moreno:2026mqk}.
\textbf{Theory automation:} symbolic and computational theory work as agent tasks.
One system enumerates 150{,}541 tree-level Feynman diagrams for multi-pair muon decay, $\mu^+\to\bar\nu_\mu\nu_e e^+ + n(e^+e^-)$, and its MadGraph cross-check of the $n=1$ branching ratio agrees with the SINDRUM measurement to within one standard deviation~\cite{Menzo:2026qrl}.
\textbf{Experiment design:} one Claude-driven design loop raised the photomultiplier count of a DarkSide-style detector from 75 to 100 in a Geant4 optical simulation, for about three times the light yield at 5~MeV.
The authors read this as a sign of a baseline poorly optimized for high energies, since the gains shrink toward the low energies that dark-matter detection cares about~\cite{Hill:2026naa}.
Cost is the next test.
\textbf{Infrastructure:} tooling that makes agents reliable.
An on-premise retrieval-augmented assistant answers paraphrased collaboration questions with a 75\% top-1 hit rate against 13\% for keyword search (the two tie on exact-keyword queries)~\cite{Mallampalli:2026hrl}.

Every number here was checked against its source paper (\Sec{sec:about}); the one hallucination that check caught, in this very cohort, is the subject of \Appx{sec:coda2}.
This is my boldest claim, and I want to report a surge rather than endorse a result: none of these papers yet publishes a measurement the community would adopt.
Open data, known answers.
Production data, unknown answers.
Closing that gap is the work of 2026--27, and it is an engineering gap: \emph{the methodology exists; the engineering rigor does not yet.}

\subsection{Claim 4: ``do we trust it?'' is the fastest-growing agenda}
\label{sec:claim4}

As deployment became routine, the field turned to trust, and I think that is healthy.
Roughly one post-split paper in five now carries a tag for uncertainty quantification, calibration, or interpretability, against one in eight in the sixteen months before the split (a share lift of 1.58).
No theme gained more papers.
Calibration is now ML: ATLAS uses an optimal-transport map (a learned deformation that carries the simulated tagger output onto the distribution seen in data) to replace per-operating-point scale factors with one continuous calibration, derived for $b$-jets~\cite{ATLAS:2025rbr}.
Theory surrogates now carry calibrated error bars.
For learned $gg\to t\bar tH$ NLO amplitudes, kernel-density resampling of the poorly learned regions and a Student's $t$ likelihood in place of a Gaussian yield uncertainties whose calibration is verified by empirical coverage~\cite{Bahl:2026qaf}.
A head-to-head comparison of repulsive ensembles, evidential regression, and Bayesian networks maps their biases and calibration limits, and all three flag numerical noise and gaps in the training data~\cite{Bahl:2025xvx}.
The category is broad: some of it produces numerical uncertainties, some calibrates classifier outputs, and some explains which physical features a tagger uses.
What unites it is that a number without a defensible uncertainty is not a measurement: \emph{the error bar is now the result.}

\subsection{Claim 5: what is slowing down is informative}
\label{sec:claim5}

I suspect some research lines shrank because they were solved or absorbed, not because interest waned.
Equivariance, the headline methods topic of 2023--24, fell from 3.3\% to 2.6\% in share.
The defining 2025 paper makes any network exactly Lorentz-equivariant, and the resulting transformers match the state of the art on JetClass tagging and set a new one on amplitude regression for $Z$ plus $n$ gluons, at $10\times$ fewer floating-point operations than L-GATr~\cite{Spinner:2025prg}, with companion work extending the construction~\cite{Favaro:2025pgz}.
The job became plumbing, and plumbing does not generate method papers.

Pure-phenomenology BSM work slipped from 13.0\% to 10.5\%, a loss roughly canceled by absolute growth in anomaly detection: the community moved from ``predict the signal'' to ``find the deviation.''
ATLAS's semivisible-jet search, for example, paired its first semi-supervised anomaly-detection region with a model-specific one, though only the latter set limits, excluding $Z'$ masses from 2 to 3.2~TeV for some invisible fractions~\cite{ATLAS:2025kuz}.
Of course, theme assignment is keyword-based and noisy, both shifts rest on small counts (17 papers to 15, and 67 to 60) and could be Poisson fluctuations alone, and the claim depends on where the line between phenomenology and anomaly detection is drawn.
The defensible statement is only that pure-phenomenology ML is shrinking while anomaly searches grow.
Either way, \emph{shrinking shares and absorbed methods are what growing up looks like.}

\subsection{Claim 6: theory ML crossed a capability threshold}
\label{sec:claim6}

A year ago, ML methods for theory were still under development.
Theory was the last frontier of ML skepticism.
It isn't anymore.
Stochastic normalizing flows combined with non-equilibrium Monte Carlo now sample topology efficiently in SU(3) Yang--Mills theory, QCD without quarks, down to a lattice spacing of $a = 0.045$~fm, where standard Monte Carlo freezes topologically and produces few or no topological-charge fluctuations~\cite{Bonanno:2025pdp}.
At the highest gluon multiplicities ($gg\to 7g$), a transformer or graph-network amplitude surrogate reaches full-color accuracy up to twice as fast as the existing chain, which generates events at leading color and reweights them to full color~\cite{Villadamigo:2025our}.
A self-supervised transformer reaches a 99.9\% solve rate on dilogarithm reductions and, with beam search, 100\% on a sample of five-point tree-level gluon amplitudes of up to 228 terms, outperforming reinforcement-learning and regression approaches~\cite{Shih:2026lmy}.
And a unified neural-network framework for nucleon imaging fits both standard lattice routes to parton distributions (quasi- and pseudo-distributions) at once, with theory constraints built into the ansatz~\cite{Chu:2025jsi}.
Admittedly, these are four demonstrations at scale in four different subfields, each on a problem whose answer was known or checkable, and none has yet displaced the tool it beat.
Still, the direction is clear: \emph{theory ML moved from fitting functions to discovering structure.}

\section{Looking ahead}
\label{sec:outlook}

Before looking ahead, here are the six claims again.
(1) The LHC experiments now publish ML-driven physics, and the ALEPH archive does too.
(2) SBI supplies the methodology and foundation models the architecture, and the two are starting to meet.
(3) Papers tagged AI agents or LLMs went from 6 to 47 and already split into four uses, with the jury still out.
(4) One recent paper in five is about uncertainty, calibration, or interpretability.
(5) Equivariance was absorbed into a tool, and model-specific phenomenology ceded ground to model-agnostic searches.
(6) Theory ML now works at scale on research areas spanning lattice topology, amplitudes, symbolic computation, and nucleon structure.

Let me separate what is settled, what is incoming, and what is still open.
\textbf{Settled:} ML is in production at the LHC experiments and in the LEP archive; SBI and unbinned measurement are the new precision standard; foundation models pretrain and fine-tune across detectors and even subfields.
\textbf{Incoming:} unbinned parton-distribution determinations, multi-experiment foundation models, and calibrated theory surrogates inside HL-LHC Monte Carlo.
Each has a proof of concept above~\cite{Barrue:2026qul, Krzmanc:2026fdw, Bahl:2026qaf}; none has a production result yet.
\textbf{Open:} whether an AI agent will publish a measurement the community trusts; the 2026 cohort makes the strongest claim of the year, but the verification cycle has not yet run.
Two other LHCP 2026 talks cover related ground: \href{https://indico.cern.ch/event/1537360/contributions/6833618/}{M.~Giacalone's plenary}, given immediately before this one, and \href{https://indico.cern.ch/event/1537360/contributions/7096100/}{T.~Lukas's parallel talk} on agents for LHC physics, reporting \Reference{Diefenbacher:2025zzn}.
The \href{https://indico.cern.ch/event/1537360/}{conference program} lists some two dozen further contributions in the anomaly-search and Performance and ML/AI sessions.

To compress three conferences into three lines: at LHCP 2025 we built the tools; at LHCP 2026 we deployed them; at LHCP 2027 we will be watching to see whether AI agents can join in.
The watching has already begun: made by an AI agent under my direction in one long weekend, this talk is itself an exhibit of the potential---and the pitfalls---of AI for the LHC physics program.

\phantomsection
\section*{Acknowledgments}
\addcontentsline{toc}{section}{\protect\numberline{}Acknowledgments}

JT thanks Siddharth Mishra-Sharma for facilitating his initial agentic AI adventures, the maintainers of the HEPML Living Review (Benjamin Nachman, Matthew Feickert, Claudius Krause, John Andrew Raine, and Ramon Winterhalder) for curating the source material for this talk, and the LHCP 2026 audience for thoughtful and impassioned engagement.
JT was supported by the U.S.\ National Science Foundation (NSF) under Cooperative Agreement PHY-2019786 (The NSF AI Institute
for Artificial Intelligence and Fundamental Interactions, \url{http://iaifi.org/}), by the U.S.\ Department of Energy (DOE) Office of High Energy Physics under grant number \mbox{DE-SC0012567}, and by the Simons Foundation through Investigator grant 929241.
JT also thanks the Institut des Hautes \'Etudes Scientifiques (IHES) and the Institut de Physique Th\'eorique (IPhT) for providing an inspiring sabbatical environment.
The AI-generated talk on which this document is based was produced by Claude Opus~4.7 in May 2026, and the bulk of these proceedings was written by Claude Fable~5.1 in September 2026 with light human editing.
The author takes full responsibility for the content of the manuscript.

\appendix

\section{Two codas after making the deck}

\subsection{Coda 1: if this talk worries you}
\label{sec:coda1}

If this talk worries you, you're right to be worried.
Some of the changes induced by the rise of AI agents have real costs, and the honest response is not reassurance but a list of unsettled issues.
On \textbf{authorship}: ``AI as research assistant'' is a credit, not authorship, and the bar---reproducible, peer-reviewed, falsifiable, justifiable---is unchanged.
On \textbf{reproducibility}: it is now a first-class methodological problem, solvable only if model versioning, archived weights, and training-data provenance are treated as part of the method.
On \textbf{commercial capture}: real, and disclosed in \Sec{sec:about}; refusing to use these tools will not fix it, whereas open weights and on-premise deployments will.
On \textbf{the next generation}: the bottleneck shifts from implementation to judgment, which is what physicists are trained for, so teach intuition harder, not less.
On \textbf{jobs and funding}: I don't know.
The field probably gets bigger before it gets smaller, because new questions become askable, but the concern is real.
On \textbf{loss of craft}: probably real.
Some of what was meaningful is being automated, and the question is what we choose to preserve deliberately.

\subsection{Coda 2: three interesting moments}
\label{sec:coda2}

At the end of the project, I asked Claude what had been most interesting about the process, and it named three moments.
The first was \textbf{a number}: the pipeline's count of papers tagged as AI agents or large language models, 6 before the split and 47 after.
Before that count, the shift was a vague impression.
After it, the shift was a load-bearing claim, and the rest of the work became pruning rather than searching.
Reading at scale buys you single numbers that organize a field.
The second was \textbf{a hallucination}: a structured reading note invented a $\tau$-polarization analysis that the agentic pipeline of \Reference{Moreno:2026mqk} never performed.
Nine reviewer passes missed it because they checked the slides against the note; the one verifier that caught it compared the slides against the source PDF.
What caught the error was not a cleverer model but the structure of the cross-checks.
The third was \textbf{a frame}: my human-authored preamble, not reproduced here, divided the subject into \emph{direct} AI/ML, which enables scientific investigations of high-dimensional spaces, and \emph{indirect} AI/ML, which capitalizes on emergent behaviors in computational systems.
Claude remarked that it could not have generated that frame, which arrived after seventeen iterations of the deck had settled on the six-claim spine.

These three moments are the exercise in miniature: what AI buys you, where it fails, and where the human contribution still lives.

\renewcommand{\bibsection}{\phantomsection\section*{\refname}\addcontentsline{toc}{section}{\protect\numberline{}\refname}}
\bibliographystyle{JHEP}
\bibliography{thaler_lhcp_2026}

\end{document}